# Free energy transduction in a chemical motor model

Josh E. Baker

Department of Molecular Physiology and Biophysics, University of Vermont, Burlington, VT 05405

To whom correspondence should be addressed: Josh E. Baker, University of Vermont, Department of Molecular Physiology and Biophysics, 115 HSRF, Burlington, VT 05405. Phone: (802) 656-3820. Fax: (802) 656-0747. E-mail: jbaker@physiology.med.uvm.edu.

**Abstract**

Motor enzymes catalyze chemical reactions, like the hydrolysis of ATP, and in the process they also perform work. Recent studies indicate that motor enzymes perform work with specific intermediate steps in their catalyzed reactions, challenging the classic view (in Brownian motor models) that work can only be performed *within biochemical states*. An alternative class of models (chemical motor models) has emerged in which motors perform work *with biochemical transitions*, but many of these models lack a solid physicochemical foundation. In this paper, I develop a self consistent framework for chemical motor models. This novel framework accommodates multiple pathways for free energy transfer, predicts rich behaviors from the simplest multi motor systems, and provides important new insights into muscle and motor function.

## Introduction

Adenosine triphosphate, or ATP, molecules are ubiquitous in cells and react with water (hydrolyze) to form the products ADP and $P_i$. Under physiological conditions, the ATP hydrolysis reaction is energetically favorable (the free energy for ATP hydrolysis, $\Delta G_{ATP}$, is negative) but slow (ATP molecules have a half-life of hours). Enzyme catalysts dramatically accelerate the rate of hydrolysis and, in the process, can harness $\Delta G_{ATP}$ to repeat a cyclical sequence of catalytic events capable of carrying out useful functions. Scheme I is a chemical representation of an enzyme-catalyzed ATP hydrolysis reaction, in which an enzyme (E) binds ATP (E.ATP), facilitates its hydrolysis by forming a stable enzyme-products complex (E.ADP.$P_i$), and then releases products in returning to its original apo state (E).

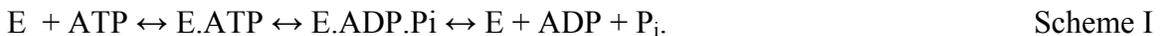

E + ATP ↔ E.ATP ↔ E.ADP.Pi ↔ E + ADP + $P_i$.         Scheme I

Molecular motors are enzymes that, over the course of their catalytic cycle, can generate force and perform work, *Fx*, in moving along a track a distance *x* against a constant force *F* (Fig. 2). Yet after decades of study, the relationship between a motor's enzymatic mechanisms and its mechanisms for force and work production remains uncertain. In 1957, A.F. Huxley developed a Brownian motor model (Huxley, 1957), later formalized by T.L. Hill (Hill, 1974), in which myosin motors in muscle *generate force* with biochemical transitions and *perform work* only within biochemical states (Fig. 1a). Similarly, in thermal ratchet models (Magnasco, 1993;Astumian and Bier, 1994) a motor *generates force* when an asymmetric potential is switched "on", presumably with a biochemical transition, and *performs work* only within the "on" state (Fig. 1b). Common to these models and to all subsequent Brownian motor models (Wang and Oster, 1998) is the assumption that a motor's translocating mechanism (in these models, a relaxation *down* a potential well) is temporally separated from a motor's enzymatic mechanism (the thermal activation of a motor *over* a potential barrier). However, it now appears that this might not be an accurate depiction of how motor enzymes really work.

Recent studies indicate that the work performed by a motor is not neatly separated from its biochemical transitions (Wang et al., 1998;Baker et al., 1999;Baker et al., 1998), and that the thermal fluctuations that activate motor enzymatic transitions might actually perform work (Fig. 1c). Based on these studies, H. Qian and others have developed chemical motor models in which motors perform work concomitant with specific biochemical steps in their catalyzed reactions (Qian, 1997;Fisher and Kolomeisky, 1999;Baker and Thomas, 2000b;Baker and Thomas, 2000a). However, most current chemical motor models lack the self-consistency found in the Brownian motor models put forth by Huxley, Hill, Magnasco, Astumian and Bier. Specifically, they violate a basic tenet of enzymology by assuming that the free energy for ATP hydrolysis, $\Delta G_{ATP}$, is diminished by the work, $Fx$, performed by the motor enzyme (Qian, 1997;Howard, 2001). This assumption is implicit in any model that describes the net free energy change around an enzyme-catalyzed reaction cycle, $RT \ln[\prod_{i=1}^{M}(k_i/k_{-i})] = \Delta G_{ATP}$, as a function of $Fx$. Here $k_i$ and $k_{-i}$ are the forward and reverse rates for step $i$ of an $M$-step ATPase reaction.

Enzymes, regardless of the forces, $F$, exerted on them, do not diminish the free energy of the reactions they catalyze, and so it follows that $Fx$ cannot be a function of $\Delta G_{ATP}$. An equivalent thermodynamic argument is that because the work performed with the ATP hydrolysis reaction is path dependent (work is only performed through the catalyzed pathway, see Fig. 2), it cannot be a state function of the hydrolysis reaction, or $\Delta G_{ATP} \neq \Delta G_{ATP}(Fx)$.

Nevertheless, $Fx$ is certainly derived from $\Delta G_{ATP}$, and describing this relationship lies at the heart of understanding free energy transduction. According to Brownian motor models, this relationship is inherently muddled because the work performed by a motor, $Fx$, is formally separated from the free energy changes associated with specific intermediate steps in the motor-catalyzed reaction (see above). Hill states, "The whole cycle is involved in the transduction process and acts as an indivisible unit" (Hill, 1989). This is not true for a chemical motor model in which the work, $Fx$, performed with a biochemical transition is derived from the free energy change for that transition. Thus to understand free energy transduction in a chemical motor model, the transfer of chemical free energy to mechanical work must be made explicit at the level of specific translocating biochemical steps in the enzyme-catalyzed reaction. Here, formally extending the model of Baker and Thomas (Baker and Thomas, 2000b), I establish a self consistent framework for free energy transduction in a chemical model of motor enzymes. This model is applicable to motor enzymes, like myosin and kinesin, that move along a track while catalyzing a hydrolysis reaction. I distinguish this model from existing motor models and develop further its implications for muscle and motor function.

**Free energy transduction through a working biochemical step.** A motor enzyme has many internal degrees of freedom, $a_1, a_2, a_3 \ldots a_N$, within its N-dimensional state space ($S_M$ in Fig. 2), where each possible combination of $a_1, a_2, a_3 \ldots a_N$ values describes a motor structural state having an energy $E(a_1, a_2, a_3 \ldots a_N)$. Ligands, such as ATP, ADP, and $P_i$, contribute to the state space of a motor ($S_M$ in Fig. 2) when they interact with the motor, but free in solution they contribute only to the state space of the ligand system ($S_{ATP}$ in Fig. 2). Figure 3 illustrates a standard, two-dimensional "energy landscape" representation of the average change in the energy, $E$, of a motor (M) as it binds and releases ligands (including its track, A) along a hypothetical motor-catalyzed ATP hydrolysis reaction pathway.

The basic assumption of a chemical motor model is that a motor enzyme performs work with a specific, reversible biochemical step in its catalyzed reaction (M.D.$P_i$ ↔ A.M.D in Fig. 3). In other words, some of the same thermally activated enzyme structural changes that facilitate the conversion of free ATP into free ADP and $P_i$ can also perform work. Specific mechanisms are discussed below (see *Motor Working Mechanisms*).

At a constant external force, the net change in a motor's energy around one reaction cycle is zero, or

$$\oint dE(a_1, a_2, a_3 \ldots a_N) = 0, \tag{1}$$

where the integral is taken around the reaction cycle through any of the innumerable pathways that exist between any two identical motor structural states. By studying these motor pathways, it is possible for structural biologists and molecular dynamicists to describe motor mechanisms (both mechanical and catalytic) in exquisite detail (McCammon and Harvey, 1987), but because a motor experiences no net energy change around its catalytic cycle (Eq. 1), this motor-centric ($S_M$) perspective provides little insight into how motors use chemical free energy to perform mechanical work.

Indeed, the work performed by a motor enzyme is not derived from changes in its internal energy, $E(a_1,a_2,a_3… a_N)$, but is instead derived from free energy changes that occur outside of a motor's state space, $S_M$, when free ATP molecules are converted into free ADP and $P_i$ molecules in $S_{ATP}$. We know this to be true for energy transduction by enzymes in general. Enzymes experience no net free energy change ($\Delta G_E = 0$) around their reaction cycle (Scheme I), but when they catalyze an energetically favorable reaction like ATP hydrolysis,

ATP ↔ ADP + $P_i$ ($\Delta G_{ATP}$ < 0 under physiological conditions),

enzymes advance unidirectionally around their cycle because the combined free energy change ($\Delta G_{ATP} + \Delta G_E = \Delta G_{ATP}$) for the coupled reaction (Scheme I) is negative.

In other words, contrary to biased diffusion models of motor enzyme function, $\Delta G_{ATP}$ does not somehow tilt the motor energy landscape ($S_M$) to make $\Delta G_E$ negative (Keller and Bustamante, 2000;Bustamante et al., 2001), nor does it introduce into the motor landscape forcing potentials that are capable of biasing motor diffusion in one direction around the reaction cycle (Mogilner et al., 2002). In fact, by focusing only on the details of motor diffusion within $S_M$ and ignoring the details of ligand diffusion and the free energy changes that occur in $S_{ATP}$, Eq. 1 implies that biased diffusion models are inaccurate representations of free energy transduction. In general, molecularly explicit models of enzyme function that do not address an enzyme's interactions with its surroundings – whether it be interactions with ligands or with ions, protons, water, etc. – are untestable, largely unconstrained, and, in the end, more nebulous than explicit.

Solution thermodynamics provides a means of imposing exact physical constraints on models of enzyme function. Within this classic chemical framework, biochemical states are coarse-grained, and biochemical transitions are described as discrete steps. This means that, in a chemical motor model, working transitions are also described as discrete working steps (Fig. 1c and see *Motor Working Mechanisms* below). In contrast, in a Brownian motor model, even when biochemical states are coarse grained, working transitions are described as continuous power "strokes" along smooth mechanical trajectories (Figs. 1a and b). As discussed below (see *Alternative Mechanisms*), power strokes are possible within the context of a chemical motor model but only as secondary working mechanisms.

Coarse-grained chemical states (and here the associated coarse-grained mechanical states) are demarcated by energy barriers in the enzyme's energy landscape, and within the intricately textured landscapes of most enzymes, the number of experimentally observable mechanical/chemical states is usually limited by the time resolution of the experimental technique used in making the observation (Frauenfelder et al., 1991). For a steady state model, mechanical/chemical states must be resolved (horizontal lines Fig. 2) for all wells within which motors equilibrate on the time scale of a steady state motor flux through the landscape.

Unlike the structural states that comprise an energy well, a mechanical/chemical state describes a time- or population-averaged motor ensemble and has a discrete energy level corresponding to the Boltzmann weighted average of all energies, $E(a_1,a_2,a_3...a_N)$, within the well it represents. Ideally we would use statistical mechanics to describe the energetics of the wells underlying chemical states, but the fact remains that we know remarkably little about these wells, not to mention how they are affected by temperature, osmotic pressure, ionic strength, pH, etc. Standard chemical potentials are used in solution thermodynamics to deal with our ignorance. The chemical potential of state $i$ is $\mu_i = \mu^o_i + RT\ln[i]$, where $\mu^o_i$ is the standard chemical potential (defined at a standard temperature, pressure, ionic strength, pH, etc.), and $[i]$ is the concentration of motors in that state (in a single motor analysis, $\mu_i = \mu^o_i + kT\ln p_i$, where $p_i$ is the probability that a motor occupies state $i$).

At fixed standard conditions, the free energy of the motor system, $S_M$, does not change when one mole of motors goes around the reaction cycle (e.g., from A.M to A.M' in Fig. 2), or

$$\Delta G_M (= \mu_{AM} - \mu_{AM'}) = 0. \tag{2}$$

This is the coarse grained chemical equivalent of Eq. 1. In contrast, when one mole of ATP molecules is hydrolyzed through either a catalyzed or uncatalyzed reaction pathway, the free energy of the ligand system, $S_{ATP}$, changes by

$$\Delta G_{ATP} = \mu_D + \mu_{Pi} - \mu_T = \Delta G^o_{ATP} + RT\ln([ADP][P_i]/[ATP]), \tag{3}$$

where $\Delta G°_{ATP}$ (= $\mu°_D + \mu°_{Pi} - \mu°_T$) is the standard reaction free energy for ATP hydrolysis. When ATP molecules hydrolyze spontaneously in solution (through $S_{ATP}$ in Fig. 2), the free energy change, $\Delta G_{ATP}$, occurs in entirety with the hydrolysis event itself; however, when ATP is hydrolyzed through an enzyme-catalyzed pathway (through $S_M$ in Fig. 2), the free energy change in $S_{ATP}$ occurs in three steps, none of which is the actual hydrolysis step.

The free energy of $S_{ATP}$ decreases by $\mu_T$ when ATP molecules bind to an enzyme, and the free energy of $S_{ATP}$ increases by $\mu_D$ or $\mu_{Pi}$ when ADP or $P_i$ molecules are released from an enzyme. Because these free energy changes occur in $S_{ATP}$, they must be harnessed at the time of ligand binding/release if they are to be used by a motor for work. A ligand chemical potential ($\mu_T$, $\mu_D$, or $\mu_{Pi}$) can be directly transferred to work if that ligand is bound/released concomitant with the working step, or a ligand chemical potential can be indirectly transferred to work by a motor if ligand binding/release induces a mechanical (see *Alternative Mechanisms* below) or chemical gradient (see below) that is subsequently used for work.

In Fig. 2, the free energy change associated with $P_i$ release, $\mu_{Pi}$, can be directly used for external work by a motor's working step, whereas the free energy changes associated with ADP release, $\mu_D$, and ATP binding, $\mu_T$, are indirectly used for external work as follows. ADP release and ATP binding effectively transfer motors from the A.M.D state to the M.D.$P_i$ state, establishing a motor concentration gradient across the working step and a negative working step free energy, $\Delta G$ (not exceeding $\mu_D - \mu_T$) that can be used for work upon the chemical relaxation of the working step. The total free energy available for work by a motor's working step is thus $\Delta\mu \equiv \Delta G + \mu_{Pi}$, and this free energy can be transferred through a motor's working step to external work, $w_{ext}$ (irreversible work, such as the net work, $Fx$, performed in moving a distance, $x$, against a dissipative force, $F$), internal work, $w_{int}$ (reversible work, such as the work performed by working steps in extending compliant elements in motors, tracks, or associated elements), or heat, $q$. Thus $n\Delta\mu = nw_{int} + w_{ext} + q$, or

$$n(\Delta\mu - w_{int}) = w_{ext} + q, \tag{4}$$

where the left side of Eq. 4 is the free energy change in the combined $S_M + S_{ATP}$ system (Fig. 2) associated with $n$ working steps, and the right side of Eq. 4 is the corresponding macroscopic work and heat lost to the surroundings ($q$ and $w_{ext}$ in Fig. 2).

Here the free energy change for the working step is diminished by $w_{int}$ not, as assumed in most current chemical motor models, by $w_{ext}$. This is because $w_{int}$ is part of the motor system (it is the work performed on the motor), whereas $w_{ext}$ is performed on something outside of the motor system. Because no net force is generated by motors in a steady state, the net internal work performed around the entire motor reaction cycle is zero, which means that $w_{int}$ performed on the motor system with the working step must be lost as heat or work elsewhere in the cycle. Thus in accord with Eqs. 2 and 3, $\Delta G_{ATP}$ is independent of both the steady state force, $F$, and the net external work, $Fx$, performed by a motor around its reaction cycle. This is clearly not the case for a model in which $Fx$ diminishes the free energy for the working step.

The time derivative of Eq. 4 gives an expression for the rate at which $\Delta\mu - w_{int}$ is transferred to external work, $w_{ext}$, and heat, $q$, by a steady state flux, $v = dn/dt$, of motors through the working step:

$$v(\Delta\mu - w_{int}) = dw_{ext}/dt + dq/dt. \tag{5}$$

The net flux, $v$, of motors through the working step can be determined from the kinetics of the entire motor-catalyzed reaction (Baker and Thomas, 2000b). According to Arrhenius kinetics, the forward and reverse working step rates depend on $w_{int}$ as

$$f_+ = f_+^\circ e^{-bw_{int}/kT} \tag{6}$$

and

$$f_- = f_-^\circ e^{-(b-1)w_{int}/kT}, \tag{7}$$

respectively, where $b$ is the fraction of $w_{int}$ performed before the transition state (Hille, 1984) and $f_+^\circ$ and $f_-^\circ$ are the forward and reverse working step rates when $w_{int} = 0$.

**Motor Working Mechanisms.** Most models of motor working mechanisms involve thermally induced changes in a motor's structure/position that are required to accommodate motor-track binding. In 1957, Huxley proposed an Eyring-like mechanism for work (Fig. 1a) in which a thermally induced extension of myosin elastic elements accommodates actin binding, and the subsequent relaxation of these elastic elements performs external work (Huxley, 1957).

The chemical model equivalent of Huxley's Eyring-like working mechanism is a Kramer-like mechanism for work (Hanggi et al., 1990) in which the thermally activated changes in a motor required for track binding directly perform external work. For example, the motor in Fig. 4 performs work when its thermal fluctuations are funneled into a rigid stereospecific motor-track complex upon track binding. The motor might be funneled as a rigid body moving toward the track (or vice versa), but the motor might also be funneled to undergo a conformational change that contributes to movement (see Fig. 4). For myosin motors, the latter model is implied both by structural studies showing a discrete rotation of the myosin motor upon actin binding (Baker et al., 1998) and by mechanical studies showing a linear relationship between the length of a myosin motor and the distance it displaces an actin track upon actin binding (Warshaw et al., 2000).

**Alternative pathways for free energy transfer.** As discussed above, ligand binding energies, $\Delta\mu$, can be transferred via a motor's working step to external work, $w_{ext}$, internal work, $w_{int}$, and heat, $q$. Here we describe how $w_{int}$ performed with a discrete working *step* mechanism can be subsequently transferred to $w_{ext}$ via a smooth power *stroke* mechanism. In the following examples we assume that a motor has an effective stiffness, $k$, and a working step of size $d$ (i.e., the average distance a motor working step moves a track against no load).

$\Delta\mu \rightarrow w_{int}$. Figure 5a shows that when the position of a track is fixed relative to a single ($N = 1$) motor, the motor performs work on itself, $w_{int} = \frac{1}{2}kd^2$, with a working step. The heat dissipated with this transition, $\Delta\mu - \frac{1}{2}kd^2$ (Eq. 5), can be reabsorbed with a working step reversal, resulting in a transfer of $w_{int}$ back to $\Delta\mu$. In other words, the working step is microscopically reversible.

$w_{int} \rightarrow w_{ext}$. Following the working step in Fig. 5a (first step), the track is allowed to move (Fig. 5a, second step) against a constant force, $F$, and the motor performs external work, $w_{ext} = F(d - F/k)$ with a power stroke mechanism. The controlled transfer of free energy from $\Delta\mu \rightarrow w_{int} \rightarrow w_{ext}$ illustrated in Fig. 5a is the assumed sequence of force-then-work generating events in Brownian motor models (Figs. 1a and 1b). In essence, Brownian motor models can be thought of as the extreme limit of a chemical model in which all of $\Delta\mu$ is transferred to $w_{int}$ (Baker et al., 2002) and none is transferred to $w_{ext}$.

Next consider a system consisting of $N = 2$ motors with a constant force, $F$, exerted on the track (Fig. 5b).

$\Delta\mu \rightarrow w_{int}$ and $w_{ext}$. Figure 5b illustrates what happens when one motor ($M_1$) undergoes its working step while a second motor ($M_2$) is already attached to the track. Motor $M_1$ performs external work, $w_{ext} = Fx_1$, in moving the track a distance $x_1$ against a force, $F$, as well as internal work both on itself, $w_{int1} = \frac{1}{2}k(\frac{1}{2}(d + F/k))^2$, and on motor $M_2$, $w_{int2} = \frac{1}{2}k(\frac{1}{2}(d - F/k))^2$. For simplicity, I have assumed that $F \ll kd$ and that the sum of the two spring displacements is $d$, which is the case if the detached motor is in register with its binding site on actin.

$w_{int} \rightarrow w_{ext}$. As illustrated in Fig. 5b (second step), after $M_2$ detaches from the track, $M_1$ performs external work, $w_{ext} = Fx_2$, with a power stroke mechanism in moving the track a distance, $x_2$, against a force, $F$, resulting in a partial transfer of $w_{int1}$ to $w_{ext}$.

The two work producing transitions (a working step followed by a power stroke) in Fig. 5b provide a novel interpretation for the observed sub-steps performed by single myosin V molecules – processive motor dimers ($N=2$ motor units) involved in intracellular transport [for review see (Mehta, 2001)]. Here I suggest that $M_1$ and $M_2$ in Fig. 5b accurately represent the lead and trail heads of myosin V, noting that in "hand-over-hand" models of myosin V, $M_2$ swings to the right of $M_1$ following its detachment from actin. According to Fig. 5b the working step of the lead head is associated with actin binding and $P_i$ release and, when $F = 0$, the working step moves an actin filament a distance $x_1 = d/2$ where $d$ has been shown to be roughly 36 nm (Mehta, 2001). A second actin displacement of comparable size ($x_2 = d/2$) is generated by a power stroke of the lead head, following the ATP-induced actin dissociation of the trail head. It can easily be shown that this symmetry between the working step and power stroke is broken if the lead head is out of register with its actin binding site.

In the one- and two-motor examples in Fig. 5, $w_{int}$ is a discrete molecular parameter (well defined for each motor) that is coupled to free energy changes of individual motors. However, in large multi-motor ensembles, it becomes virtually impossible to localize $w_{int}$ to individual motors, and mechanochemical coupling (i.e., the relationship between $w_{int}$ and $\Delta\mu$) is most easily described by a continuous, macroscopic $w_{int}$ that is coupled to the molar free energy changes of the bulk motor system. In active, isometric muscle it has been shown that $w_{int} = \bar{F}d$, where $\bar{F}$ is the molar muscle force (Baker and Thomas, 2000a).

In a chemical model of a bulk motor system, the exchange of free energy between two continuous forms ($\Delta\mu \leftrightarrow w_{int}$) makes mechanochemical oscillations a distinct possibility. I begin by considering the mechanical response of a bulk motor system to a force pulse. When a collection of motors reaches a stall force and is then mechanically perturbed by $\delta = \Delta \bar{F} d$ through a force jump, the motor system responds with a chemical relaxation of the working step and a corresponding force response (Fig. 6a). Assuming that $f_+ = f_-$ at stall (Baker and Thomas, 2000b), the initial recovery rate is $r = f_+ + f_- = A(\exp[-b\Delta \bar{F} d/RT] + \exp[-(1-b)\Delta \bar{F} d/RT])$ (Eqs. 6 and 7). This equation (with b = 0) resembles the equation used by Huxley and Simmons (Huxley and Simmons, 1971) to describe the initial rate of force recovery following a rapid step in the length of an isometric muscle, only here the operative mechanical parameter is a macroscopic force not, as in the Huxley-Simmons equation, a molecular strain.

In many muscle types, the complete force response of muscle to a rapid length or force step resembles that of a damped harmonic oscillator (time and frequency domains plotted in Figs. 6d and e). This response is most often described as a sum of multiple processes [phases 2 and 3 in the time domain and processes "C" and "B" in the frequency domain (Kawai and Brandt, 1980)]. However, it may be that motor ensembles in muscle behave as resonant systems. The energetic requirements for resonance are met in this model by an exchange of free energy between $w_{int}$ and $\Delta \mu$, and a dynamic instability of the motor ensemble provides a possible kinetic mechanism for resonance (see Fig. 6). Further experimental and theoretical studies will be needed to test this novel hypothesis, but its implications for oscillatory motor systems, like the heart and insect flight muscle, are immediately clear. For example, it may be that the heart functions optimally when its mechanics and kinetics are tuned for resonance and that the loss of mechanochemical resonance, as well as compensatory attempts to regain it, might be a basic, yet unexplored, cause of certain heart diseases.

DISCUSSION

Over the past twenty years, major technological advances have enabled us to directly measure the relationship between motor enzyme structure, mechanics, and chemistry in single molecules and in bulk motor systems. Many of these studies have challenged conventional perspectives on how motors work (Brownian motor models in Figs. 1a and b) and have led to the emergence of an alternative class of motor models referred to as chemical motor models. Here I discuss some of the fundamental differences between Brownian and chemical motor models and summarize the basic physicochemical requirements for chemical motor models, established in this paper.

The first Brownian motor model was proposed by A.F. Huxley (Huxley, 1957), formalized by T.L. Hill (Hill, 1974), and subsequently adapted in thermal ratchet models (Astumian and Bier, 1994). The basic features of this model include:

1. Work and force production. A motor's translocating mechanisms are formally separated from its enzymatic mechanisms. That is a motor's reaction and position, $x$, coordinates are assumed to be orthogonal (see Figs. 1a and b). A motor generates force with a thermally-activated biochemical transition (often described as a coarse-grained discrete step) and performs work only subsequently with a power stroke (a smooth change in $x$ in Figs. 1a and b).

2. Mechanochemical coupling. Reaction free energies are defined as a function of a motor's position coordinate, $x$, with the assumption being that $x$ remains constant during a biochemical transition (i.e., $x$ is not a reaction coordinate). Indeed Brownian motor models are uniquely characterized by x-dependent reaction free energies and reaction rates.

3. Free energy transduction. The transfer of chemical free energy to mechanical work is not localized to a specific biochemical step in the motor catalyzed reaction (see introduction).

Recent studies have challenged the Brownian motor paradigm, indicating that motors perform work with relatively discrete translocating steps that are closely associated with biochemical steps. In accord with these observations, numerous chemical motor models have been proposed (Qian, 1997;Fisher and Kolomeisky, 1999;Baker and Thomas, 2000a;Howard, 2001). However, most current chemical motor models erroneously assume an enzyme-dependent $\Delta G_{ATP}$. In this paper, I have developed a more self-consistent framework for chemical motor models. The basic features of this model include:

1. Work and force production. A motor's translocating mechanisms and enzymatic mechanisms are inseparable and must be treated on equal footing (i.e., $x$ need not be constant during a biochemical transition; Fig. 1c). A motor can generate both force and work with a thermally activated biochemical transition (Fig. 3). In a coarse grained chemical motor model, a motor working transition, like its associated chemical transition, is described as a discrete step (not a power stroke) and is presumably driven by a motor conformational change that is induced by ligand binding/release (Fig. 4). In a chemical motor model, power strokes can occur as secondary working mechanisms (Fig. 5) and contribute to rich behaviors from the simplest multi motor systems (see *alternative pathways* above).

2. Mechanochemical coupling. In this paper, I have argued that the free energy available for work by the working step, $\Delta\mu$, is diminished by the internal work, $w_{int}$, performed with the working step not, as assumed in most chemical motor models, by the external work, $w_{ext}$ (Eq. 4). In a steady state, the net internal work performed by a working motor around its catalyzed reaction cycle is zero and thus neither $w_{int}$ nor $w_{ext}$ diminish $\Delta G_{ATP}$ in accord with Eqs. 2 and 3.

3. Free energy transduction. $\Delta\mu$ can be transferred to external work, $w_{ext}$, either directly through a working step mechanism or indirectly through a power stroke mechanism ($w_{int} \rightarrow w_{ext}$). In either case, the transfer of chemical free energy to mechanical work can be explicitly traced back to the motor working step.

A third class of motor enzyme models, referred to as biased diffusion models, assumes that motor diffusion is biased by a net tilt or driving potential across the motor landscape (Keller and Bustamante, 2000;Mogilner et al., 2002). This model might be applicable to motor systems that modify (tilt) their landscape during translocation, like DNA repair enzymes. However, the potential gradient that drives translocation by the class of motors discussed in this paper (i.e., motors that catalyze a hydrolysis reaction as they move along a track) is a potential gradient in the ligand system ($\Delta G_{ATP}$ in $S_{ATP}$) not in the motor system ($S_M$). Thus, as discussed in this paper, biased diffusion models do not apply to these motors.

Before we can claim to understand how molecular motors really work, we must develop explicit models for both force and work generation by motors in relation to a motor's enzymatic mechanisms; we must legitimize these models using basic physicochemical principles; and we must experimentally test these models to determine which ones most accurately describe motor enzyme function. This paper's contribution is to legitimize the chemical motor hypothesis and explicitly distinguish it from Brownian and other motor models.

ACKNOWLEDGEMENTS

I thank the members of the Department of Molecular Physiology and Biophysics at the University of Vermont, with special thanks to D. Warshaw, J. Moore, N. Kad, and G. Kennedy for their insight and guidance.

FIGURE LEGENDS

Figure 1. Motor models. (a) According to the Huxley-Hill model, reaction (ordinate) and position (abscissa) coordinates are orthogonal. Motor force (the slope of a potential well) is generated with a motor biochemical step (down arrow) and work is subsequently performed (curved arrow) when a motor relaxes within the potential well of a biochemical state. (b) Similarly, according to thermal ratchet models, reaction (ordinate) and position (abscissa) coordinates are orthogonal. Motor force (the slope of a potential well) is generated when a ratchet potential is switched on, and work is subsequently performed when a motor relaxes within the potential well of the "on" state. (c) In a chemical motor model, reaction and position coordinates are intimately linked. Force is generated and/or work is performed with a thermally activated biochemical transition. In all three models, the biochemistry is coarse-grained and discrete.

Figure 2: Pathways for ATP hydrolysis and energy transfer. In system $S_{ATP}$, held at constant temperature and pressure, ATP molecules hydrolyze to form the products ADP and $P_i$. ATP hydrolysis can occur in solution (straight arrow in $S_{ATP}$) or through an enzyme-catalyzed pathway (curved arrow through the motor enzyme system, $S_M$). Regardless of the pathway, when one mole of ATP molecules is hydrolyzed, the free energy of $S_{ATP}$ changes by $\Delta G_{ATP}$. In $S_M$, motor enzymes (double ovals) move a track (helix) against a force, $F$, as they catalyzed the ATP hydrolysis reaction, generating external work, $w_{ext}$, and heat, $q$.

Figure 3: A motor energy landscape. The curved line represents hypothetical changes in a motor's average energy along its catalyzed ATP hydrolysis reaction pathway. The straight lines represent chemical potentials of coarse grained motor chemical states. The landscape excludes chemical potentials of ligands. A common feature of many motor enzymes is that their track affinity is modulated along their enzymatic reaction pathways. Here the modulating biochemistry resembles that of the actin-myosin catalyzed ATP hydrolysis reaction. Briefly, ATP (T) binds to myosin (M), inducing the dissociation of myosin from actin (A). ATP bound to myosin (M.T) is then hydrolyzed (M.D.$P_i$). Actin accelerates the release of $P_i$, and work is performed (arbitrarily illustrated by a turning wheel against a force, $F$) upon actin-binding and $P_i$ release (M.D.$P_i$ to A.M.D). Myosin returns to its original state (A.M) with the release of ADP.

Figure 4. A possible mechanism for a motor's working step. To structurally accommodate binding to a track (black notched rectangle) a motor (scissors-like structure) undergoes an internal structural change, resulting in a displacement, $x$, of the track. For myosin and actin, the biochemistry of the motor-track binding transition is the M.D.$P_i$ to A.M.D transition.

Figure 5. Multiple pathways for energy transfer in one- and two-motor systems. (a) A single motor can perform internal work, $w_{int}$, in extending an elastic element (spring) upon track binding if the track is held at a fixed position (first step). (b) This internal work can subsequently be transferred to external work, $w_{ext} = Fx$, if the track is allowed to move a distance, $x$, against a load, $F$ (second step). (c) If one motor ($M_1$) undergoes its working step (M.D.P$_i$ → A.M.D) while a second motor (M$_2$) is already bound to a track, M$_1$ performs $w_{int}$ on both M$_1$ and M$_2$ as well as $w_{ext} = Fx_1$ in moving the track a distance, $x_1$, against a load, $F$ (first step). When M$_2$ subsequently detaches from the track, the internal work that was performed on M$_1$ is transferred to $w_{ext} = Fx_2$ in moving the track a distance, $x_2$, against a load, $F$ (second step).

Figure 6. Mechanochemical oscillations. (a) At a steady state stall force, $\Delta\mu = w_{int}$ (Baker and Thomas, 2000a). (b) When $w_{int}$ is rapidly increased by $\delta$ through a jump in the force exerted on a track, motors respond with a net working step reversal, a decrease in $w_{int}$ (i.e., force), and a corresponding increase in $\Delta\mu$. The opposite response is elicited following a rapid decrease in $w_{int}$. (c) If the motors in (b) undergo a mechanically concerted working step reversal, cooperatively preventing forward working steps from occurring, motors cannot redistribute for maximal entropy, and the free energy perturbation $\delta$ is transferred to $\Delta\mu_L$. The motor system will oscillate until $\delta$ is dissipated as heat or the motors are able to redistribute for maximal entropy. (d) A polar plot of the response of a damped harmonic oscillator, having a characteristic frequency of 170 sec$^{-1}$ (comparable to the working step relaxation rate of skeletal muscle myosin) and a damping coefficient of 100 (black), 70 (dark gray), and 50 (light gray). (e) The response of the three oscillators in (d) to a force step.

Figure 1

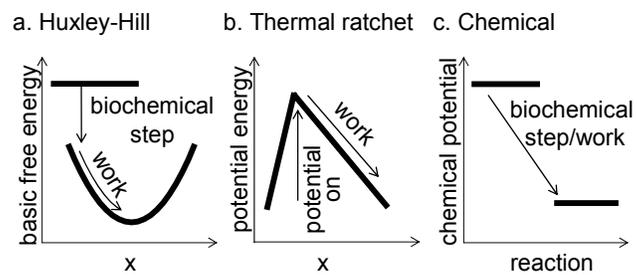

Figure 2

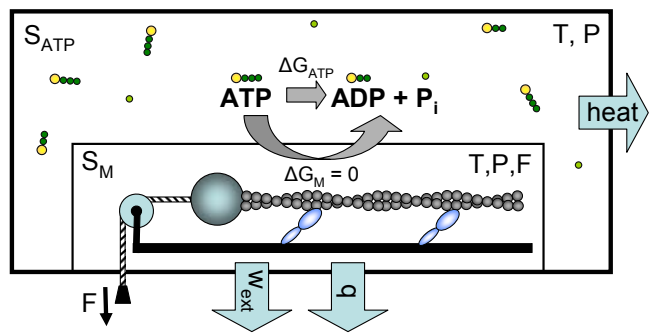

Figure 3

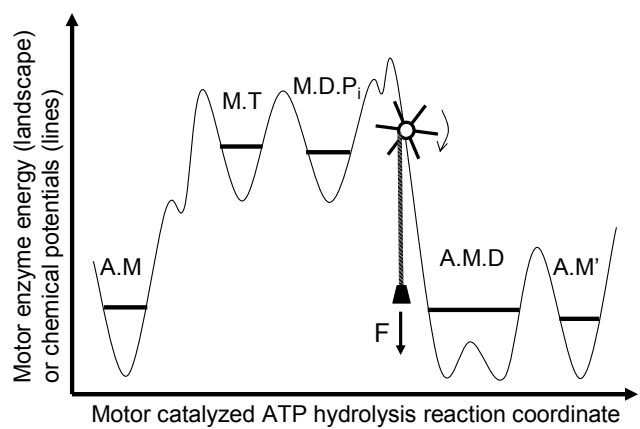

Figure 4

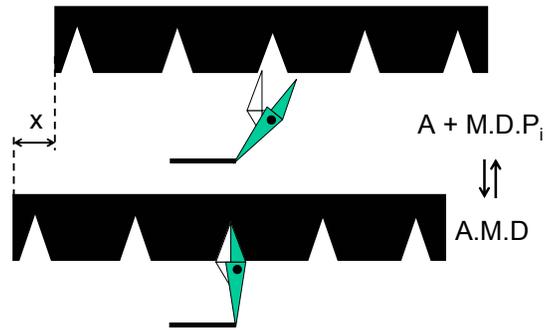

Figure 5

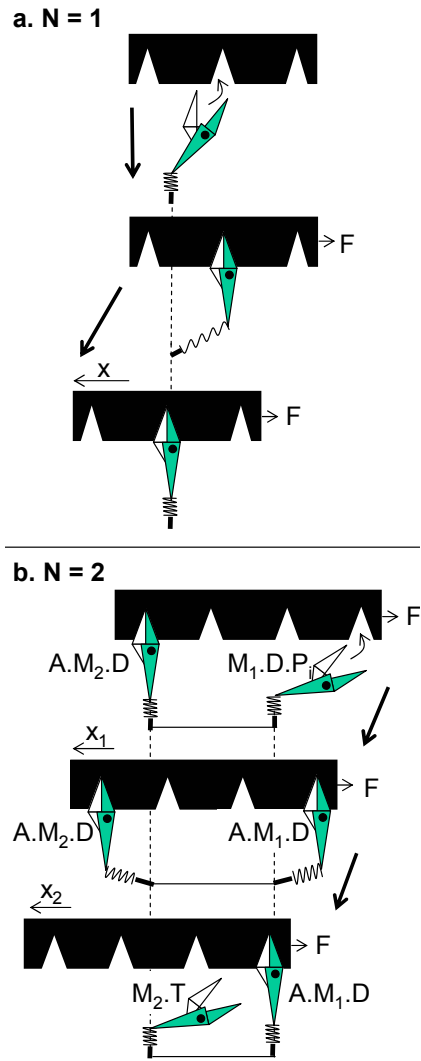

Figure 6

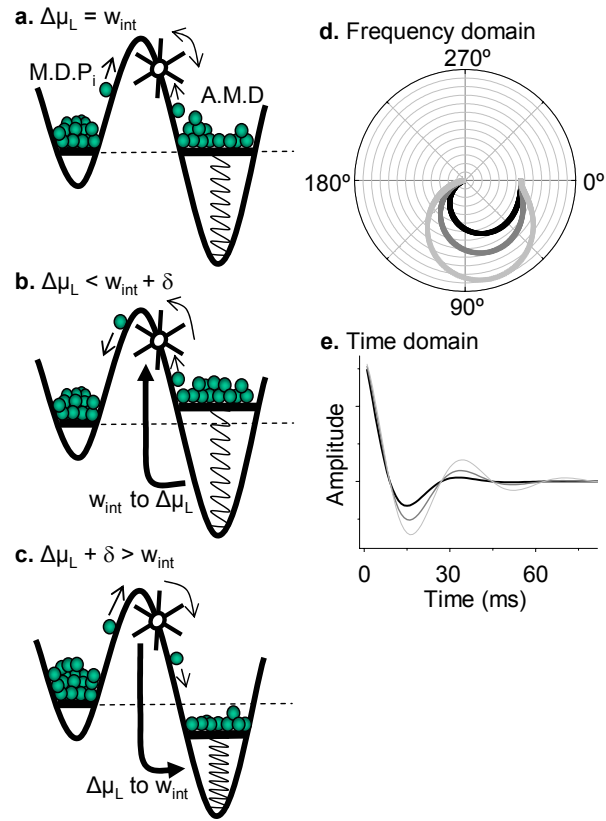